\documentclass[twocolumn,preprintnumbers,amsmath,amssymb,nofootinbib]{revtex4}
\usepackage{graphicx}
\usepackage{bm}
\usepackage{latexsym}
\usepackage{epsfig}
\usepackage{natbib}
\usepackage{color}

\begin{document}

\title{Matter-Antimatter Asymmetry Induced by a Running Vacuum Coupling}

\author{J. A. S. Lima}
\email{jas.lima@iag.usp.br}
\affiliation{Departamento de Astronomia, Universidade de S\~ao Paulo, Rua do Mat\~ao 1226, 05508-900, S\~ao Paulo, Brazil}

\author{D. Singleton}
\email{dougs@csufresno.edu}
\affiliation{Department of Physics, California State University Fresno, Fresno, CA 93740-8031, USA \\
and \\
Institute of Experimental and Theoretical Physics Al-Farabi KazNU, Almaty, 050040, Kazakhstan}

\begin{abstract}
We show that a CP-violating interaction induced by a derivative coupling between the running vacuum and a non-conserving baryon current may dynamically break CPT and  trigger baryogenesis through an effective chemical potential. By assuming a nonsingular class of running vacuum cosmologies which provides a complete cosmic history (from an early inflationary de Sitter stage to the present day quasi-de Sitter acceleration), it is found that an acceptable baryon-asymmetry is generated for many different choices of the model parameters. It is interesting that the same ingredient (running vacuum energy density) addresses several open cosmological questions/problems: avoids the initial  singularity, provides a smooth exit for primordial inflation, alleviates both the coincidence and cosmological constant problems,  and, finally,  is also capable of explaining the generation of matter-antimatter asymmetry in the very early Universe. 

\end{abstract}

\date{\today }
\maketitle

\hskip 0.1cm {\it Introduction.--} There is a growing body of work on running vacuum cosmologies or an effective  dynamical $\Lambda$-term \cite{Marek15,AS2015,Pigozzo2016,Sola2016,Lima2016}. These models are motivated by the cosmological constant problem (CCP) -- the absence of a satisfactory mechanism whereby the vacuum energy density from all fields can be canceled (or almost canceled) by the fixed cosmological constant. In other words, we do not know how the current, extremely small value of the effective vacuum energy density can be predicted from first principles.

Although successful, the current cosmic concordance model ($\Lambda_0$CDM + Inflation) can also be seen as a collection of many different ingredients that are brought together to explain the complete cosmic evolution over all time and length scales. Such disparate ingredients are needed in order to smoothly connect the early and late  time accelerating regimes. Beyond the enigmatic CCP \cite{Wein89,Sahni00}, there are also specific  challenges for the cosmic concordance model: the initial singularity, the `graceful' exit problem for some popular models of inflation, cosmic coincidence \cite{Stein97} (see also \cite{RP88,RHB1,PRL2,JCAP2016} for these and other potential problems).  

The interest in running vacuum cosmologies can be justified both phenomenologically and from first principles based on quantum field theory in curved spacetimes. The geometric $\lambda$-term (proportional to the metric and usually present on the {\it l.h.s.} of Einstein's equations) can be incorporated on the {\it r.h.s.} of Einstein's equations as part of the effective energy momentum tensor (EMT) involving also additional vacuum contributions of the matter sector, $T_{\mu \nu} = \Lambda_{E}M_{Pl}^{2}\,g_{\mu\nu}$, where $M_{Pl}=(8\pi G)^{-1/2}\sim 2.4\times 10^{18}$ GeV is the reduced Planck mass and $\Lambda_{E}$ is the effective $\Lambda$-term.  {The question arises as to what is the total expression of $\Lambda_{E}$? The vacuum state of all existing fields can be represented by an EMT which reflects the Lorentz invariance of its energy density and pressure.  By averaging over all fields the minimally coupled EMT in general relativity  reads  $\langle T_{\mu \nu} \rangle  = \langle \rho \rangle g_{\mu \nu}$. When   the geometric $\lambda$-term  is transferred to the {\it r.h.s.} of Einstein's equations one may see that both contributions add yielding \cite{Wein89,Sahni00}

\begin{equation}
\label{le}
\Lambda_{E}= \lambda + \frac{\langle \rho \rangle }{M_{Pl}^{2}}\,, 
\end{equation}
with the field equations becoming $G^{\mu \nu} = M_{Pl}^{-2}[{T^{\mu\nu}_{(f)}} + \Lambda_{E}M_{Pl}^{2}g^{\mu \nu}]$. Since the Einstein tensor is divergenceless  ($\nabla_{\mu} G^{\mu \nu} = 0$), it is usually argued (even in textbooks) that the vacuum energy density remains constant ({\it i.e.} $\nabla_{\mu} \Lambda_{E}\equiv \partial_{\mu} \Lambda_{E} = 0$). However,  this result holds only if the cosmic fluid EMT is always separately conserved ($\nabla_{\mu} {T^{\mu\nu}_{(f)}}=0$). In the general case one finds: 
\begin{equation}
\nabla_{\mu} G^{\mu \nu} = 0 \, \Rightarrow \nabla_{\mu}\, {T^{\mu\nu}_{(f)}} = - g^{\mu \nu}\nabla_{\mu} \Lambda_{E}\,,
\end{equation}  
which implies that a running  $\Lambda_{E}$  can transfer energy to the cosmic fluid ($T^{\mu\nu}_{(f)}$).  In particular, for spatially inhomogeneous and time-dependent space-times} even a local dependence of the effective cosmological term, $\Lambda_{E}\equiv \Lambda_{E}(x^{\mu})$, is possible because energy transfer at different points of the spacetime may occur at different rates.

In cosmology, the assumption  of a running vacuum energy density is also physically more appealing than the standard view of a fixed $\Lambda_{E}$-term. Since the physical quantities at the background level depend only on time, the condition  $\partial_{0} \Lambda_{E} \neq 0$ means that the extremely low value currently observed  ($\Lambda_{E}(t_0)\equiv\Lambda_0$) can be theoretically  accommodated. For an aged Universe,  the present smallness of $\Lambda_{E}(t)$ can be seen as a natural consequence of its oldness.

This contemporary view has observational support. Recent works on a class of running vacuum models have shown that a battery of tests including {\it SNIa, BAO, H(z), LSS, BBN, CMB}  provide a quality fit that is significantly better than the $\Lambda_0$CDM \cite{AS2015,Sola2016}. Several authors, following different lines of inquiry, have investigated how the running vacuum transfers energy to the material components. Probably, the most successful  path comes from QFT techniques  based on the renormalization group (RG) approach in curved spacetimes. Broadly speaking, one may think that the vacuum energy density `runs' because the effective action inherits quantum effects from the matter sector. Generically, the RG techniques in curved spacetimes leads to a dependence $\rho_v(H,\dot H) = M_{Pl}^{2}\Lambda_E(H,\dot H)$ where H is the Hubble parameter (see \cite{PLB000,Sola2016} and references there in).

Given the possibility of a running vacuum cosmology, it is also essential to investigate how the {\it baryogenesis problem}, {\it i.e.} the observed matter-(anti)matter asymmetry, can be handled in this context. This old problem becomes more intriguing  in light of the novel running vacuum scenario where the space-time emerges from a pure nonsingular de Sitter stage smoothly evolving into radiation, matter and $\Lambda_0CDM$, thereby avoiding the initial singularity and providing a complete description of the expansion history \cite{Lima2013,P2013}. One may ask whether the baryogenesis mystery can also be explained by the same running vacuum driving the cosmic dynamics. 

In this work we  address this challenging problem by proposing a new mechanism based on a derivative CP-violating coupling between the running vacuum and the baryon number current. As we shall see, the predicted matter-(anti)matter asymmetry is in good agreement with the observations for many different choices of the relevant cosmic energy scales.

{\it Running vacuum baryogenesis.--} Currently, the baryon asymmetry (B-asymmetry for short) is numerically characterized by the dimensionless  $\eta$-parameter:
\begin{equation}
\label{eta}
5.7\times 10^{-10} < \eta  \equiv \frac{n_B - n_{\bar B}}{s}< 6.7 \times 10^{-10}\,,
\end{equation}
where $n_{B}$, $n_{\bar B}$ are  the number densities of baryons (anti)-baryons,  respectively, and $s$ is the present radiation entropy density. The lower and upper limits on the $\eta$-parameter come from precision measurements of the primordial deuterium abundance and cosmic background radiation (CMB) acoustic peaks \cite{eta,Planck1}. The exact nature of the  baryogenesis mechanism is not known yet. Some  models have been proposed to explain the B-asymmetry,  but there is no consensus in the literature concerning  the correct approach \cite{spontB,LI02,Brand03,GL2007,steinhardt} (see also \cite{review,Rev2016} for reviews). 

We now  discuss a new mechanism appropriate for running $\Lambda_E(t)$-models (to simplify notation, $\Lambda_{E} (t) \equiv \Lambda(t)$). The basic ingredient is a  derivative coupling between the running vacuum and the baryon current $J^{\mu}_{B}$: 
\begin{equation}\label{action}
S=\frac{1}{M_{\Lambda}^{2}}\int {d^4}x{\sqrt{-g}}(\partial_\mu\Lambda)J^{\mu}_B\,,
\end{equation}
where $M_{\Lambda}$ is an unknown cutoff mass scale of the effective decaying vacuum theory whose natural upper bound is the reduced Planck mass $M_{Pl}$. Like the effective $\Lambda$-term from \eqref{le}, the coupling \eqref{action} can viewed either as a particle physics motivated mechanism as in the spontaneous baryogenesis model of \cite{spontB} which has a coupling of the form 
\begin{equation}\label{action-sb}
S=\frac{1}{M_{SB}}\int {d^4}x{\sqrt{-g}}(\partial_\mu \phi)J^{\mu}_B\,,
\end{equation}
(where $M_{SB}$ is the cutoff mass scale and $\phi$ is a scalar field)
or a space-time inspired mechanism, as in the gravitational baryogenesis model of \cite{steinhardt} which has a coupling of the form
\begin{equation}\label{action-gb}
S_{GB}=\frac{1}{M_{*}^{2}}\int {d^4}x{\sqrt{-g}}(\partial_\mu R)J^{\mu}_B\,,
\end{equation}
(where $M_{*}$ is the cutoff mass scale and $R$ is the Ricci scalar). The proposed coupling in \eqref{action} is a mixture of spontaneous baryogenesis and gravitational baryogenesis, keeping the good features of each without some of the drawbacks that both the spontaneous baryogenesis and gravitational baryogenesis mechanisms have. As in the gravitational baryogenesis scenario, the multiplicative cutoff mass scale, $M_{\Lambda}$, in \eqref{action} has a mass dimension of [$\rm mass]^{-2}$  instead of [$\rm mass]^{-1}$  as in the Cohen-Kaplan spontaneous baryogenesis model \cite{spontB} or similar scenarios \cite{LI02,Brand03,GL2007}. The coupling given in \eqref{action} shares, with the gravitational baryogenesis model, several advantages with respect to the spontaneous baryogenesis model: (i) the scalar field of spontaneous baryogenesis must be assumed to evolve homogeneously in the spatial directions in order to produce the baryon asymmetry while in the present model and that of \cite{steinhardt} the homogeneity is naturally built into the model; (ii) in the spontaneous baryogenesis model the oscillations of the scalar field tend to time average to zero which washes out the baryon asymmetry. 

The coupling in \eqref{action}, though different, has advantages to the gravitational baryogenesis model as given by the coupling in \eqref{action-gb}. Taking into account the spatial homogeneity of the FRW metric one finds that in \eqref{action-gb} $\partial _\mu R \rightarrow {\dot R} = 0$ at tree-level. Thus at tree-level the gravitational baryogenesis  mechanism does not produce a baryon asymmetry. This is handled in \cite{steinhardt} by arguing that to the one-loop level in an SU(N) gauge theory which some fixed number of flavors of quarks one finds that ${\dot R} \ne 0$. However this assumes that nothing else comes in to these one-loop corrections as one goes to higher energy scales. In the model given by the coupling in \eqref{action} one automatically gets baryogenesis as soon as one considers a non-constant $\Lambda$ without the need to consider higher order loop corrections as in \cite{steinhardt}.     

By requiring a B-violation process in such  thermal quasi-equilibrium state  it is easy to show that an acceptable B-asymmetry can be easily generated. This happens  because the above interaction also implies that CPT is dynamically broken thereby driving the energetics to cause the B-asymmetry. The process is actually very similar to gravitational baryogenesis \cite{steinhardt} which relies  on the derivative coupling between the Ricci curvature scalar $R$ and the B-current. However, the running vacuum process proposed here has two new interesting features:  (i) the same ingredient driving the early accelerating phase of the Universe ({\it i.e.} the running vacuum) may also control the baryogenesis process; (ii) the running vacuum is always accompanied by particle production and entropy generation \cite{Waga93,Lima96,Lima2016}. This nonisentropic process is an extra source of T-violation (beyond the freeze out of the B-operator) which as first emphasized by Sakharov \cite{Shaka67} is a basic ingredient  of baryogenesis.
 
As remarked before, all quantities, including the running $\Lambda$, are only time-dependent (or more accurately
depend on the standard FRW time parameter and time foliation). Hence,  in the effective Lagrangian \eqref{action},  one may replace $\partial_\mu  \Lambda \rightarrow \partial_0 \Lambda \equiv{\dot \Lambda} <0$, thereby giving rise to an effective  chemical potential.  This inequality sign is required by the second law of thermodynamics when the vacuum decays into massless particles in the very early Universe \cite{Waga93,Lima96}. For a species of particle $i$ carrying a baryon charge of $q_i$  we can define an effective  chemical potential as 
\begin{equation}
\label{chem-pot}
\mu _{Bi} = -q_i\frac{\dot \Lambda}{M^2 _{\Lambda}} = 
\pm \frac{{\dot\Lambda}}{M^2_{\Lambda}}\,,
\end{equation}
where in the last step was assumed that all baryons-(anti)baryons are characterized by baryon numbers $\pm 1$, respectively. The  B-asymmetry produced by the above chemical potential can be written as:
\begin{equation}
\label{eta-2}
\eta = \frac{n_B}{s} \approx -\frac{{\dot \Lambda}}{M_{\Lambda}^{2} T} {\Big \rvert} _{T=T_D}\,,
\end{equation}
with the temperature evaluated at $T=T_D$ when the B-violation mechanism decouples. To obtain the above expression one
needs the quasi-equilibrium relationship for the entropy per unit volume when the vacuum decays ``adiabatically'' \cite{Lima96}, $s = \frac{\rho + p}{T} \propto T^3$. Replacing $\dot \Lambda$ by $\dot R$, where $R$ is the Ricci scalar,  \eqref{eta-2} becomes exactly the  prediction of the gravitational baryogenesis process \cite{steinhardt,LD2016}.  In \eqref{eta-2} we have approximated the multiplicative factor, $15g_b/4\pi^{2}g_* \sim {\cal O} (1)$, where $g_b$  ($g_*$)  counts the baryonic (all) degrees of freedom.

The dependence of the $\eta$-parameter with the rate of the running $\Lambda$, as given by \eqref{eta-2},  is a key result of this work. In order to illustrate its consequences, we will estimate the generation of the B-asymmetry for a  class of nonsingular running vacuum cosmology recently proposed \cite{AS2015,Lima2013,P2013}.

Such running vacuum cosmologies describe a complete cosmological history evolving from an early to a late time de Sitter state. The vacuum energy density is characterized by a  truncated power-series in the Hubble parameter, $H \equiv {{\dot a}}/{a}$, in which its dominant term behaves like $\rho_{\Lambda}(H) \equiv M^{2}_{Pl}\Lambda (H) \,\propto H^{n+2}, n > 0$. Such models have some interesting features, among them: (i) regardless the value of $n$, the universe starts from a nonsingular de Sitter phase ($H=H_I$) and evolves smoothly  to the radiation phase, giving a successful model for a graceful exit, (ii) the temperature and entropy of radiation is initially zero (de Sitter vacuum), but the entropy growth due to running $\Lambda(H)$ generates all the present day observed entropy, and  the radiation temperature evolution law has also been analytically determined \cite{Lima2016,Lima15}, (iii) the late-time expansion history is a realistic competitor with the rigid $\Lambda_0$CDM model \cite{AS2015,Lima2016}. 
  
The total energy conservation law in running vacuum cosmologies, 
$u_{\alpha}{T^{\alpha \beta}_{(f)}}_{;\beta}= -u_{\alpha}{T^{\alpha \beta}_{(\Lambda)}}_{;\beta}$,  reads:
\begin{equation}\label{ECL}
 \dot\rho + 3H (\rho + p) = - M^{2}_{Pl} {\dot \Lambda}\,,
\end{equation}
where $\rho$ and $p$ are the energy density and pressure of the fluid component. It shows that in the expanding Universe, the variation of $\Lambda$ is only possible with particle and energy transfer from the running vacuum to the fluid component (see \cite{Lima96} for a general treatment). One can see immediately that the above equation can also be directly obtained by combining the cosmological equations with variable $\Lambda$ \cite{CLW92,LM94}. For simplicity, we consider here a subclass in which the late time observed flat Universe is exactly the concordance $\Lambda_0$CDM model while at early times one can have a varying cosmological constant. This kind of model is captured by the phenomenological expression
\begin{equation}\label{L}
{\Lambda(H)} = \Lambda_0 + 3 \nu H^2 + 3 H^{2}\left(\frac{H}{H_I}\right)^{n}\,.
\end{equation}
where the power index $n>0$, $\nu$ is some parameter, $H_I$ is an arbitrary inflationary scale and the factors of 3 are introduced for mathematical convenience. Such models where proposed in \cite{GCL2015,Lima2013,P2013,Lima2016} and were initially purely phenomenologically motivated to give a early inflationary phase ({\it i.e.} the last term $\propto H^{2}\left(\frac{H}{H_I}\right)^{n}$), a late time constant $\Lambda _0$ term, and an intermediate term  $\propto \nu H^{2}$. Later models havig a varying $\Lambda$ of the form given in \eqref{L} were more rigorously motivated by applying the renormalization group approach from quantum field theory in curved spacetimes \cite{PLB000}.
Note that for $H<<H_I$ the model behaves like the $\Lambda_0$CDM cosmology. From observations over the redshift interval $0 \leq z \leq 1100$ \cite{AS2015,Sola2016} one finds that the parameter from the quadratic term, $3 \nu H^2$, is $\nu \sim {\cal O} (10^{-3})$. Thus at the early times considered here, and where baryogenesis occurs, this term is negligible \newpage 
\noindent and we will set $\nu =0$ in the rest of the paper \footnote{The main effect of $\nu$ in  the solution \eqref{Ha} for $H(a)$ is to change the energy scale $H_I$  by an effective scale, $\tilde H_I=(1-\nu)^{1/n}H_I$, and the power $n$ is replaced by $\tilde n=n(1-\nu)$ (see Eq. (3.6) in Ref. \cite{Lima2016} and subsequent discussion). For the current accepted $\nu$ values such corrections are small at the baryogenesis epoch and have been neglected in the present analysis. Naturally, this does not means that such a quadratic $\nu$-term is negligible at late times (radiation and vacuum-matter stages).}.

Taking the two Friedmann equations ({\it i.e.} the $00$ component equation $3 H^2 = 8\pi G \rho + \Lambda(H)$ and the $ii$ component equation $-2\dot H - 3H^{2} = 8\pi G p - \Lambda(H)$), combining this with $\Lambda (H)$ from \eqref{L} and taking the limit that one is at early times so that $\Lambda _0$ can be neglected and so that $\omega = \frac{p}{\rho} = \frac{1}{3}$ one obtains the evolution equation for the Hubble parameter of the form \cite{GCL2015}:
\begin{equation}\label{EM1}
 \dot H + {{2}}H^{2}\left(1- \frac{H^n}{H_I^n}\right)=0\,,
\end{equation}
showing that for $\dot H=0$ one has $H=H_I$ which represents the  primordial de Sitter stage. This phase is unstable, and the  general analytical solution reads: 
\begin{equation}\label{Ha}
H (a) =\frac{H_{I}}{\left[1+({a}/{a_{eq}})^{2n}\right]^{1/n}}\,, 
\end{equation}
where $a_{eq}$ is the vacuum-radiation equilibrium value of the scale factor, that is, when  $\rho_{\Lambda} = \rho_{rad}$, a moment coincident  with the end of inflation  ($\ddot a = 0, H (a_{eq})\equiv H_{end} = H_I/2^{1/n}$). For $a << a_{eq}$ we find $H=H_I$ (de Sitter) while for $a >> a_{eq}$ the solution reduces to the standard radiation phase, $a(t) \propto t^{1/2}.$ Therefore, the solution \eqref{Ha}  describes a smooth transition from a primeval non-singular Sitter stage to the standard FRW phase regardless of the power index $n$. This result points to some universality of the process and also suggests a natural ``graceful" exit from the early unstable de Sitter stage to the standard radiation epoch when the particle production ends \cite{Lima2016, LM94, GCL2015}.  
Under certain conditions, the running vacuum may decay into massless particles preserving some equilibrium relations.  This happens when the vacuum decays adiabatically. Physically, this means that radiation entropy is generated, effectively massless particles-(anti)particles are  created, but the specific entropy per particle remains constant. In this case, it has also been shown that $\rho_{rad} \propto T_{rad}^{4}$ and $n_{rad} \propto T^{3}$ \cite{Lima96}. Further, by  combining \eqref{L} and the first Friedmann equation, $\rho_{rad} + M^{2}_P\Lambda (H) = 3M^{2}_PH^{2}$,  and taking $({\frac{90}{\pi^2 g_*}})^{1/4} \sim 10^{-1/4}$, we obtain the radiation temperature law as a function of $H$:
\begin{equation}\label{HFT}
T_{rad}(H)= 2^{\frac{n+2}{4n}}T_{end}\left(\frac{H}{H_I}\right)^{1/2}\left[1 - \left(\frac{H}{H_I}\right)^n\right]^{\frac{1}{4}}\,,
\end{equation}
where $T_{end}= T(H_{end})=10^{-\frac{1}{4}}\sqrt{M_{Pl}H_I}$ is the temperature at the end of inflation. Note also that $T_{rad}=0$ for $H=H_I$.  Hence, there is no thermal bath in the begin of the cosmological evolution. 

{\it Baryogenesis constraints.--} In order to obtain the strength of the B-asymmetry we need to calculate $\dot \Lambda$ and the temperature $T=T_D$ when the B-violation mechanism decouples (see Eq. \eqref{eta-2}). Using \eqref{L} and \eqref{EM1} we obtain $\dot \Lambda$:  
\begin{equation}
\label{t-dot-R-2}
{\dot \Lambda } = -6(n+2)H^3\left(\frac{H}{H_I}\right)^{n} 
\left[ 1- \frac{H^n}{H_I^n} \right]\,.
\end{equation}
To calculate the decoupling temperature $T_D$ we need to specify the B-violating operator to determine when the B-violating process decouples. To simplify the discussion we take a 
generic GUT, B-violating operator having mass dimension 6 {\it e.g.} ${\cal L}_{\Delta B} = \frac{1}{M_X ^2} \epsilon ^{\alpha \beta \gamma} \epsilon^{ab} ({\bar u}^c _\gamma \gamma^\mu q_{\beta a})
({\bar d}^c _\alpha \gamma_\mu l_b)$ where $u, d, q, l$ stand for up-quark, down-quark, quark and lepton respectively. The rate of this generic dimension 6 interaction is  
$\Gamma _{\Delta B} \sim \frac{T^5}{M_X ^4}$ with $M_X$ being the mass scale connected with ${\cal L}_{\Delta B}$. The decoupling temperature is fixed by requiring 
$\frac{T_D ^5}{M_X ^4} \sim \Gamma _{\Delta B} \le H \approx \frac{T_D ^2}{M_{Pl}}$ (the expression for $H$ comes from \eqref{HFT} assuming $H < H_I$ so that the last term is of order 1; also we have taken $2^{-(n+2) / 2n} \sqrt{10}$ of order 1).    
Using this we find the decoupling temperature and decoupling $H$ as 
\begin{equation}\label{HD}
T_D \sim \frac{M_X ^{4/3}}{M_{Pl} ^{1/3}} ~~~~{\rm and} ~~~~ H_D \sim \frac{M_X ^{8/3}}{M_{Pl} ^{5/3}} \,.
\end{equation} 
Using $H_D$ and $T_D$ and Eqs. \eqref{t-dot-R-2} and \eqref{eta-2} we find 
\begin{equation}
\label{etaD2} 
\eta \approx 6(n+2) \left( \frac{M_{Pl} ^2}{M_\Lambda ^2} \right) \left( \frac{H_I}{M_{Pl}} \right) ^{(5n+20)/3} \left( 1 - \frac{H_I ^{5n/3}  }{M_{Pl} ^{5n/3} }  \right) ~.
\end{equation}
For simplicity we have assumed that the inflation scale and B-violating operator mass scale are of the same order -- $H_I \sim M_X$. This is in accord with some simple models of inflation where the inflation scale and GUT scale coincide. 

The general predictions for $\eta$ depend on the values assigned to $n$, $M_{Pl}/M_{\Lambda}$, and $H_I/M_{Pl}$. These parameters should be chosen to give the observed $\eta$ and be in agreement with the inflationary scale. In Table I, we display the running vacuum predictions of the B-asymmetry  for a broad set of selected  values of the three free parameters. Note as
the index $n$ increase $M_{Pl}/M_{\Lambda}$ increases if we keep $H_I/M_{Pl}$ in a range that gives $H_I \sim 10^{15} - 10^{16}$ GeV {\it i.e.}the GUT scale which is then consistent with our simplifying assumption of $H_I \sim M_X$. 
The main conclusion from Table I is that one can obtain values of $\eta$ in running vacuum models of the kind discussed here for a range of the relevant sub-Planckian parameters. Thus, this framework provides a successful  baryogenesis mechanism driven by a running vacuum model as described here. Finally we note that in all cases $H_I < M_{Pl}$ thereby showing that trans-Planckian problems are absent.

\vskip 0.3cm
\begin{table}[h!]
\centering
\begin{tabular}{|c|c|c|c|c|}
\hline  $n > 0$   & \ $M_{Pl}/M_{\Lambda} \geq 1$ & \ $H_I/M_{Pl}\leq 1$  \ & \ $\eta$      
\\
\hline  $0.1$ & $3.0\times 10^5$ & $8.2 \times 10^{-4}$   \ &  \ $6.4 \times 10^{-10}$ 
\\  
\hline $1.0$ & $3.4 \times 10^5 $  & \ $2.6\times 10^{-3}$ & \ $6.0 \times 10^{-10}$ 
\\
\hline $2.0$ & \ $8.2 \times 10^4 $ & \ $9.0 \times 10^{-3}$  & \  $5.6 \times 10^{-10}$ 
\\
\hline $3.0$ & \ $2.6 \times 10^6 $ & \ $9.6 \times 10^{-3}$  & \  $5.8 \times 10^{-10}$ 
\\
\hline $4.0$ & \ $5.4 \times 10^8 $ & \ $7.6 \times 10^{-3}$  & \ $5.8 \times 10^{-10}$ 
\\  
\hline $5.0$ & \ $7.1 \times 10^9 $ & \ $9.1 \times 10^{-3}$  & \  $6.1 \times 10^{-10}$ 
\\
\hline $6.0$ & \ $8.9 \times 10^{11}$ & \ $8.2\times 10^{-3}$  & \  $6.5 \times 10^{-10}$ 
\\
\hline
\end{tabular}
\caption{Baryogenesis prediction in a class of running vacuum cosmologies. Note that the above values, which are in good agreement with the current constraints on $\eta$ -- see equation \eqref{eta} -- and the inflationary scale \cite{Planck}, are obtained for a range of the model parameters.}
\end{table}

{\it Final remarks.--} We have investigated the early generation of B-asymmetry driven by a nonsingular running vacuum cosmology. It should be stressed that the B-violation as discussed here does not require new ingredients. This happens because the running  vacuum is the same quantity driving the evolution of the space-time (see Eqs. \eqref{action}, \eqref{eta-2}, \eqref{L} and \eqref{EM1}). Interestingly, the adopted running vacuum model is also endowed with some  remarkable properties  like the avoidance of the initial singularity and a smooth exit from inflation to the standard radiation phase thereby producing all the observed entropy. It also alleviates the cosmic coincidence and $\Lambda$ problems, and,  last but not least, we have found the its decay products, through the derivative coupling with the $\Lambda(t)$ term given in \eqref{action}, is also  capable of generating the observed baryon asymmetry.

It is worth noticing that the new mechanism proposed here is different from spontaneous baryogenesis as discussed by several authors \cite{spontB,LI02,Brand03,GL2007}, as well as from  gravitationally induced B-asymmetry powered by the Ricci scalar \cite{steinhardt,LD2016,Pan},  a Gauss-Bonnet term \cite{GBonnet} or other higher order curvature invariants. Probably, the closest approach is the gravitational baryogenesis driven by the Ricci scalar in the presence of a running vacuum \cite{Pan}. However, even in this case there are significant differences. For example, the Ricci scalar is $R = -{M_{Pl}^{-2}} T - 4\Lambda$, where $T\equiv \rho -3p$ is the trace of the matter energy-momentum tensor. For relativistic massless particles (radiation phase, $T\equiv 0$),  one finds $\dot R= - 4\dot \Lambda$. Accordingly, gravitational baryogenesis will predict a value of $\eta$ four times larger than in the present  approach. This occurs because the effective chemical potential (and $\eta$) in gravitational baryogenesis are proportional to ${\dot R}$ \cite{steinhardt} while for the direct running vacuum coupling these quantities are proportional to ${\dot \Lambda}$ (see Eqs. \eqref{chem-pot}-\eqref{eta-2}). In addition, for $T \ne 0$, or equivalently, if the EoS parameter is slightly different from $\omega = 1/3$, as discussed in Ref.\cite {steinhardt}, then gravitational baryogenesis will have two distinct positive contributions associated with the pair of physical quantities ($\dot T, \dot \Lambda$). This means that the result $\eta_{GB} = 4\eta_{\Lambda}$ is a lower bound.  

As can be seen from Table I, the cut-off mass scale scale of the effective theory ($M_{\Lambda}$) does not need to be tuned to Planck mass in order to generate the observed value  $\frac{n_B}{s} \sim 10^{-10}$. 

Finally, we stress that running vacuum models are able to address a broader class of cosmological observations (including baryogenesis) within a single mechanism \cite{Lima2016,GCL2015}. In the case of the cosmological constant problem, for instance, one may compare the vacuum energy densities in the early de Sitter regime and the present day value. By taking the constraints of $H_I$ from Table I, we see that such a ratio satisfies the inequality $\rho_{vI}/\rho_{v0}\simeq\Lambda_I/\Lambda_0 \geq 10^{108}$.

{\par\noindent {\bf Acknowledgments:}} JASL was partially supported by CNPq, CAPES (PROCAD 2013) and FAPESP (Brazilian Research Agencies). DS is supported  by grant $\Phi.0755$  in fundamental research in natural sciences by the Ministry of Education and Science of Kazakhstan.


\begin{thebibliography}{99}

\bibitem{Marek15} M. Szydlowski, Phys. Rev. D {\bf91}, 123538 (2015), arXiv:1507.02114

\bibitem{AS2015} A. Gomez-Valent and J. Sol\`a , Mon. Not. R. Astron. Soc. 448  (2015) 2810; {\it ibid.}, Astrophys. J. {\bf 811}, L14 (2015).

\bibitem{Sola2016}  J. Sol\`a, ``Running Vacuum in the Universe: current phenomenological status'', arXiv:1601.01668

\bibitem{Pigozzo2016} C. Pigoso {\it et al.}, JCAP {\bf 022}, 1605 (2016) no.05, 022, arXiv:1510.01794  

\bibitem{Lima2016}  J. A. S. Lima, S. Basilakos and J. Sol\`a, Eur. Phys. J. {C76}, 228 (2016), arXiv: 1509.00163

\bibitem{Wein89} S. Weinberg. Rev. Mod. Phys. {\bf 61}, 1 (1989).

\bibitem{Sahni00} V. Sahni and A. Starobinsky, IJMP D {\bf 9}, 373 (2000).

\bibitem{Stein97} J. P. Steinhardt, in ``Critical Problems of Cosmology'', Eds. V. L. Fitch, D. R. Marlow, M. A. E. Dementi, Princeton UP 
(1997).

\bibitem{RP88} B. Ratra and P. J. E. Peebles, Phys. Rev. D {\bf 37}, 3406 (1988).

\bibitem{RHB1} R. H. Brandenberger, Braz. J. of Physics {\bf 31}, 131 (2001). {\it ibid.},  {\it A Status Review of Inflationary Cosmology}, arXiv:hep-ph/0101119

\bibitem{JCAP2016}  A. Del Popolo, J. A. S. Lima, J. C. Fabris, D. C. Rodrigues, JCAP 1404 (2014) 021, arXiv:1404.3674

\bibitem{PRL2} F. C. Carvalho, J.S. Alcaniz, J.A.S. Lima and R. Silva, Phys. Rev. Lett. {\bf 97},  081301 (2006), astro-ph/0608439

\bibitem{PLB000} J. Sol\`a, J. Phys. Conf. Ser. {\bf 453}, 012015 (2013); I. L. Shapiro and J. Sol\`a, Phys. Lett. B 475, 236 (2000).

\bibitem{Lima2013} J. A. S. Lima, S. Basilakos and J. Sol\`a, MNRAS {\bf 431}, 923 (2013), arXiv:1209.2802 

\bibitem{P2013} E. L. D. Perico, J. A. S. Lima, S. Basilakos and J. Sol\`a, Phys. Rev. D{\bf 88}, 063531 (2013), arXiv:1309.0591 

\bibitem{eta} K.A. Olive {\it et al.} (Particle Data Group), Chin. Phys. C, {\bf 38}, 090001 (2014); R. Y. Cooke, {\it et al.}  Astrophys. J. 781, 31 (2014).

\bibitem{Planck1} P. A. R. Ade, N. Aghanim, M. Arnaud et al., “Planck 2015 results. XIII. Cosmological parameters,” http://arXiv.org/abs/
1502.01589.

\bibitem{spontB} A. G. Cohen and D. B. Kaplan, Phys. Lett. B {\bf 199}, 251 (1987).

\bibitem{LI02} M.-z. Li, X.-l. Wang, B. Feng, and X.-m. Zhang, Phys. Rev. D {\bf 65} (2002) 103511, [hep-ph/0112069].

\bibitem{Brand03}  R. H. Brandenberguer and M. Yamaguchi, Phys. Rev. D {\bf 68}, 023502 (2003).

\bibitem{GL2007} G. Barenboim and J. D. Lykken, JHEP {\bf 10} (2007) 032, [arXiv:0707.3999].

\bibitem{steinhardt} H. Davoudiasl, R. Kitano, G. D. Kribs, H. Murayama and P. J. Steinhardt, Phys. Rev. Lett. {\bf 93}, 201301 (2004); arXiv:hep-ph/0403019.

\bibitem{review} A. Riotto and M. Trodden, Annu. Rev. Nucl. Part. Sci. {\bf 49}, 35 (1999); M. Dine and A. Kusenko, Rev. Mod. Phys. {\bf 76}, 1 (2003).

\bibitem{Rev2016}  E. Mavramatos, J. of Physics: Conference Series 447, 012016 (2013).  

\bibitem{Shaka67} A. D. Shakarov, JETP Lett. {\bf 5}, 24 (1967).

\bibitem{Waga93} J. M. Salim and I. Waga, Class. Quant. Grav. {\bf 10}, 1767 (1993). See also \cite{Lima96} for a general thermodynamical analysis  of running vacuum models. 

\bibitem{Lima96} J. A. S. Lima, Phys. Rev. D {\bf 54}, 2517 (1996), gr-qc/9605055 

\bibitem{LD2016} J. A. S. Lima and D. Singleton, Phys. Lett. B {\bf 762}, 506 (2016), arXiv:1610.01591 

\bibitem{Lima15} J. A. S. Lima, S. Basilakos  and J.  Sol\'a,  Gen. Rel. Grav. {\bf 47} (2015) 40, arXiv:1412.5196 

\bibitem{CLW92} J. C. Carvalho, J. A. S. Lima, Phys. Rev. D {\bf 46}, 2404 (1992); I. Waga, Astrophys. J. {\bf 414}, 436 (1993).

\bibitem{LM94} J. A. S. Lima and J. M. F. Maia,  Phys. Rev. D {\bf 49}, 5597 (1994).

\bibitem{GCL2015} J. A. S. Lima, G. J. M. Zillioti and R. C. Santos, arXiv:1508.06344 
 
\bibitem{Planck} Planck Collaboration, Planck 2015 results. XX. Constraints on inflation", arXiv:1502.02114.

\bibitem{Pan} V. K. Oikonomou, S. Pan and R. C. Nunes, arXiv:1610.01453v1

\bibitem{GBonnet} S. D. Odintsov and V. K. Oikonomou, arXiv:1607.00545v1


\end{thebibliography}
\end{document}